\documentclass[aps,prl,superscriptaddress,amsfonts,amsmath,amssymb,twocolumn,showpacs,floatfix,nofootinbib]{revtex4-1}

\usepackage{amsmath}
\usepackage{amssymb}
\usepackage{graphicx}

\newcommand{\beq}{\begin{equation}}
\newcommand{\eeq}{\end{equation}}
\newcommand{\bma}{\begin{math}}
\newcommand{\ema}{\end{math}}
\newcommand{\beqa}{\begin{eqnarray}}
\newcommand{\eeqa}{\end{eqnarray}}

\newcommand{\id}{{\mathbf 1}}

\allowdisplaybreaks[4]

\newcommand{\be}{\begin{equation}}
\newcommand{\ee}{\end{equation}}
\newcommand{\bq}{\begin{eqnarray}}
\newcommand{\eq}{\end{eqnarray}}

\newcommand{\rf}[1]{(\ref{#1})}

\begin{document}

\title{Condensate-induced transitions and critical spin chains}

\author{Teresia M\r{a}nsson}
\affiliation{%
Department of Theoretical Physics, School of Engineering Sciences,
Royal Institute of Technology (KTH),
Roslagstullsbacken 21, SE-106 91 Stockholm, Sweden
}

\author{Ville Lahtinen}
\affiliation{%
Nordita, Royal Institute of Technology and Stockholm University,
Roslagstullsbacken 23,
SE-106 91 Stockholm,
Sweden
}
\affiliation{%
Institute for Theoretical Physics, University of Amsterdam,
Science Park 904,
1090 GL Amsterdam,
The Netherlands
}

\author{Juha Suorsa}
\affiliation{%
Nordita, Royal Institute of Technology and Stockholm University,
Roslagstullsbacken 23,
SE-106 91 Stockholm,
Sweden
}

\author{Eddy Ardonne}
\affiliation{%
Nordita, Royal Institute of Technology and Stockholm University,
Roslagstullsbacken 23,
SE-106 91 Stockholm,
Sweden
}
\affiliation{%
Department of Physics, Stockholm University,
AlbaNova University Center, SE-106 91 Stockholm, Sweden
}

\date{\today}
\begin{abstract}
We show that condensate-induced transitions between two-dimensional topological phases provide a general framework to relate
one-dimensional spin models at their critical points. We demonstrate this using two examples. First, we show that two well-known spin chains, namely the XY chain and the transverse field Ising chain with only next-nearest-neighbor interactions, differ at their critical points only by a non-local boundary term and can be related via an exact mapping. The boundary term constrains the set of possible boundary conditions of the transverse field Ising chain, reducing the number of primary fields in the conformal field theory that describes its critical behavior. We argue that the reduction of the field content is equivalent to the confinement of a set of primary fields, in precise analogy to the confinement of quasiparticles resulting from a condensation of a boson in a topological phase. As the second example we show that when a similar confining boundary term is applied to the XY chain with only next-nearest-neighbor interactions, the resulting system can be mapped to a local spin chain with the $u(1)_2 \times u(1)_2$ critical behavior predicted by the condensation framework.
\end{abstract}


\maketitle

An essential characteristic of topologically ordered phases is the absence of local order parameters \cite{wen95}.
While this property ensures a degree of stability, which is of potential benefit in quantum information technologies, it also makes it necessary to develop new approaches for the study of phase transitions in these systems. As the low-energy degrees of freedom are anyonic quasiparticles specific to a particular topological phase, it should be possible to understand transitions between the phases in terms of their collective behavior. One such successful framework is that of condensate-induced transitions \cite{bais09}, where the condensation of a bosonic particle induces confinement of a set of quasiparticles, thereby causing a transition to a new phase. 
Topological phase transitions in various systems, including the quantum Hall hierarchies \cite{halperin83, haldane83, suorsa11, bonderson12, barkeshli10}, interacting anyons \cite{gils09, ludwig11, lahtinen12} and Levin-Wen models
\cite{levin05,burnell11}, can be related to this mechanism.
It is well known that 2+1-dimensional topological phases are intimately related to
two-dimensional conformal field theories (CFT) \cite{witten89, moore89}: The topological charges carried by 
the gapped bulk quasiparticles are in one-to-one correspondence with the primary fields of a CFT describing the gapless edges.
It is then natural to ask whether the condensate-induced transitions have a counterpart in critical one-dimensional systems that are also described by CFT's.

In this Letter, we answer this question positively and show that the framework of condensate-induced transitions relates also critical one-dimensional spin chains. We demonstrate this using two examples. First we revisit the relation between the XY chain \cite{lsm61} and two copies of the transverse field Ising (TFI) chain \cite{epw70,p70}. The criticality of these chains is described by the $u(1)_4$ and the Ising $\times$ Ising CFT's, respectively, which are related by the condensation framework \cite{bais09} when viewed as low-energy theories of topological phases.
Going beyond the previous works that considered open boundary conditions \cite{jf78,pjp82,f94}, we show that for closed boundary conditions the two spin chains differ at their critical points only by a term that constrains the set of possible boundary conditions for the TFI chains. We argue that these constraints, in effect, realize the confinement of those CFT primary fields that correspond to those bulk quasiparticles confined following a bulk condensation. To demonstrate the generality of this picture, we apply a similar confining boundary term to two copies of the XY chain and derive a microscopic spin chain with the predicted critical $u(1)_2 \times u(1)_2$ behavior. Condensate-induced transitions \cite{bais09} offer thus a powerful general framework for both relating known critical spin chains as well as deriving novel ones with a given CFT description. Due to connections between the ground states of critical spin chains and gapped topological phases \cite{Thomale10}, critical spin chains may thus offer a simple setting to study complex topological phase transitions.

\paragraph{The transverse field Ising model --}
The critical transverse field Ising chains that we consider are given by the Hamiltonian
\begin{equation}
H_{\rm TFI}^n = \sum_{i=0}^{L-1}  
\sigma^{z}_{i} + \sigma^{x}_{i} \sigma^{x}_{i+n} \ ,
\label{eq:tfi-nn}
\end{equation}
where we assume periodic boundary conditions, $\sigma^{\alpha}_L = \sigma^{\alpha}_{0}$. The critical nearest-neighbor chain ($n=1$) can be solved via a Jordan-Wigner transformation, which
transforms the Pauli operators $\sigma^{\alpha}_i$ into fermionic operators $c^{\dagger}_i$ and
$c^{\vphantom\dagger}_i$ \cite{epw70,p70}. The operators $c^{\dagger}_i$ involve strings of
Pauli operators, which we choose to start at site zero and end at site $i$.
The resulting Hamiltonian,
\begin{align}
H^{1}_{\rm TFI} =&
\sum_{i=0}^{L-1} (2c^{\dagger}_{i} c^{\vphantom{\dagger}}_{i}-1) +
\sum_{i=0}^{L-2}
(c^{\vphantom{\dagger}}_{i} - c^{\dagger}_{i}) (c^{\vphantom{\dagger}}_{i+1} + c^{\dagger}_{i+1})
\nonumber \\
&-\mathcal{P} (c^{\vphantom{\dagger}}_{L-1} - c^{\dagger}_{L-1})
(c^{\vphantom{\dagger}}_{0} + c^{\dagger}_{0})\, .
\label{eq:fermion-tfi}
\end{align}
conserves fermionic parity, which is described by the symmetry operator $\mathcal{P} = \prod_{i=0}^{L-1} \sigma^{z}_i = \exp(i\pi \sum_i c_i^\dagger c_i)$. The boundary conditions for the fermions are thus in one-to-one correspondence with parity sectors:  for odd parity ($\mathcal{P}=-1$) one has $c_{L} \equiv c_{0}$, while for even parity ($\mathcal{P}=1$), one has $c_{L} \equiv -c_{0}$. In momentum space the Hamiltonian takes the form (after a Bogoliubov transformation) 
\begin{align} 
H_{\rm TFI}^1 &= -4 \sum_{k}
\bigl|
\sin\bigl( \frac{\pi k}{L}\bigr)
\bigr|
\bigl( c^{\dagger}_{k} c^{\vphantom{\dagger}}_{k} -\frac{1}{2} \bigr)  \ , 
\label{eq:tfi-nn-momentum}
\end{align}
where the $c^{\dagger}_{k}$ create fermions with momentum $k$. Due to the parity-dependent boundary conditions, these momenta take
integer values for $\mathcal{P}=-1$ and half-integer values for $\mathcal{P}=1$.

The Ising CFT (with central charge $c=1/2$) describes the the criticality of the TFI chain. 
%The CFT describing the criticality of the TFI chain, i.e. whose spectrum matches the low energy spectrum of the system, is the Ising
%CFT with central charge $c=1/2$.
This CFT has three primary fields, $1$, $\sigma$ and $\psi$, with scaling dimensions
$h_1 = 0$, $h_\sigma = 1/16$ and $h_\psi = 1/2$. These fields label the states of the energy spectrum of the TFI chain such that those in the even parity sector ($\mathcal{P} = 1$, anti-periodic boundary conditions) are labeled by either $1$ or $\psi$, while all the states the
odd parity sector ($\mathcal{P} = -1$, periodic boundary conditions) are labeled by $\sigma$.  We refer to \cite{byb} for more details on CFT.

Here we are interested in the TFI chain with only next-nearest-neighbor interactions ($n=2$). The Hamiltonian $H_{\rm TFI}^2$ decouples into two TFI chains on the even (e) and the odd (o) sites, each of which can be solved in the same way as the nearest neighbor TFI chain above. The fermion parity is now conserved independently for the even and odd sites, with the corresponding symmetry operators given by $\mathcal{P}_{e} = \prod_{j} \sigma^{z}_{2j}$ and
$\mathcal{P}_{o} = \prod_{j} \sigma^{z}_{2j+1}$. The CFT describing the critical behavior is the direct product of two Ising CFT's (denoted as
${\rm Ising}^2$), with total central charge $c=1$. The correspondence between the symmetry sectors of $H_{\rm TFI}^2$ and the primary fields of the ${\rm Ising}^2$ CFT is shown in Table \ref{tab:sectors}. 
\begin{table}[th]
\begin{tabular}{c}
Sectors of $H^2_{\rm TFI}$\\
\begin{tabular}{|c|c|}
\hline
 $(BC_e,BC_o)$  & ${\rm Ising}^2$ fields \\
	\hline
$(1,1)$ & $(1,1)$,$(1,\psi)$,$(\psi,1)$,$(\psi,\psi)$  \\
\hline
$(1,-1)$ & $(1,\sigma)$, $(\psi,\sigma)$ \\
\hline
$(-1,1)$ & $(\sigma,1)$, $(\sigma,\psi)$ \\
\hline
$(-1,-1)$ & $(\sigma,\sigma)$  \\
	 \hline
\end{tabular}
\end{tabular}
\begin{tabular}{c}
Sectors of $H^1_{\rm XY}$\\
\begin{tabular}{|c|c|}
\hline
$\mathcal{T}^{z}$   & $u(1)_4$ fields \\
	\hline
$1$ &  $\tilde{1}$, $\tilde{\psi}$ \\
\hline
$-1$ & $\lambda$, $\bar{\lambda}$\\
\hline
\end{tabular}
\end{tabular}
\caption{\label{tab:sectors}%
{\it Left:} The correspondence between the symmetry sectors, that are in one-to-one correspondence with the boundary conditions $(BC_e,BC_o)=$($\mathcal{P}_e$,$\mathcal{P}_o$) of $H^{2}_{\rm TFI}$, and the nine primary fields of the ${\rm Ising}^2$ CFT that label the states in them.
{\it Right:} The correspondence between the symmetry sectors of $H^{1}_{\rm XY}$ and the four primary fields of the $u(1)_4$ CFT.}
\end{table}

\paragraph{The XY spin chain --}  The second spin chain we consider is the critical XY chain with periodic boundary conditions. Its Hamiltonian is given by 
\begin{equation}
H_{\rm XY}^n = \sum_{i=0}^{L-1} \tau^{x}_{i} \tau^{x}_{i+n} + \tau^{y}_{i} \tau^{y}_{i+n} \ ,
\label{eq:XY}
\end{equation}
where $\tau_i^{\alpha}$ are Pauli matrices and, as above, $n=1$ ($n=2$) corresponds to a model with only nearest-neighbor (next-nearest-neighbor) interactions. Also the XY chain can be diagonalized with a Jordan-Wigner transformation \cite{lsm61}, with the fermion parity now corresponding to the symmetry operator $\mathcal{T}^z = \prod_i \tau^{z}$. Like in \rf{eq:fermion-tfi}, it will again determine the boundary conditions for the fermions, and thus the allowed momenta in the momentum space where $H_{\rm XY}^1$ is diagonal:
\begin{align}
H_{\rm XY}^1 &= 4 \sum_{k} \cos \bigl( \frac{2 \pi k} {L} \bigr)
\bigl( c^{\dagger}_{k} c^{\vphantom{\dagger}}_{k} -\frac{1}{2} \bigr) \ .
\label{eq:XY-momentum}
\end{align}
We assume that $L$ is even for which case the XY chain has further structure: The operators $\mathcal{T}^{x} = \prod_{i=0}^{L-1} \tau^{x}_i$ and
$\mathcal{T}^{y} = \prod_{i=0}^{L-1} \tau^{y}_i$ commute with Eq.~\eqref{eq:XY}, with each other and with $\mathcal{T}^{z}$. The sectors $\mathcal{T}^{z} = \pm 1$ each thus split into two sectors, which are degenerate in the case $\mathcal{T}^{z} = -1$.

The XY chain at the critical point can be described in terms of the $u(1)_4$ CFT of a chiral boson \cite{byb}. 
This CFT has central charge $c=1$ and contains four primary fields $\tilde{1}$, $\lambda$, $\bar{\lambda}$ and $\tilde{\psi}$ with
scaling dimensions $h_{\tilde{1}} = 0$, $h_{\lambda} = h_{\tilde{\lambda}} = 1/8$ and $h_{\tilde{\psi}} = 1/2$. The labeling of the states follows again from the boundary conditions in each parity sector. As illustrated in Table~\ref{tab:sectors}, states in the even parity sector ($\mathcal{T}^z=1$, anti-periodic boundary conditions) are associated with the vacuum $\tilde{1}$ or the fermion $\tilde{\psi}$ primary fields, while the degenerate states in the odd parity sector ($\mathcal{T}^z=-1$, periodic boundary conditions)  are associated with the pair of primary fields $\lambda$ and $\bar{\lambda}$.

We will show below that when a boundary term $H^{B}_{\rm TFI}$ is introduced,
$H^2_{\rm TFI}+ H^{B}_{\rm TFI}$ can be mapped exactly to $H^1_{\rm XY}$. To assign a physical meaning to the boundary term, we set the spin chains briefly aside and review the notion of condensate-induced transitions between gapped topological phases \cite{bais09}.

\paragraph{The condensation framework --}
In topological field theories,  condensate-induced transitions proceed in three steps that we outline in the context of the ${\rm Ising}^2$ theory. The labels $1$, $\psi$ and $\sigma$ of an Ising theory should be understood as topological quantum numbers of the quasiparticle excitations with the following `tensor product', or fusion rules:
\be \label{Ising_fusion}
	\psi \times \sigma = \sigma, \quad \psi \times \psi = 1, \quad \sigma \times \sigma = 1 + \psi\,.
\ee
Fusion is symmetric ($a \times b = b \times a$), distributive ($(a+b) \times c = a \times c + b \times c $), associative ($a \times (b\times c) = (a \times b) \times c$) and all particles fuse trivially with the vacuum label ($a \times 1 = a$). The fusion rules of the ${\rm Ising}^2$ theory follow directly from the fusion rules of each Ising theory.

In our example, the first step of condensation is to identify the boson $(\psi,\psi)$ with the vacuum label $(1,1)$. By definition, the vacuum should have trivial fusion and statistics with all the other particles. Thus, in order that $(\psi,\psi)$ behaves like the vacuum, the first step is to identify all particles $a$ and $b$ that are related by fusion with it. That is, if $a \times (\psi,\psi) = b$, then we set $a = b$.
We arrive at a new set of particle types $\tilde{1} = (1,1) = (\psi,\psi)$,
$\tilde{\psi}=(1,\psi)=(\psi,1)$,
$\tilde{\sigma}_1=(\sigma,1)=(\sigma,\psi)$ and
$\tilde{\sigma}_2=(1,\sigma)=(\psi,\sigma)$,
while the particle $(\sigma,\sigma)$ remains unaffected at this step. In the second step, one demands that all particles should have trivial statistics with the new vacuum. This is equivalent to demanding that identified particles with unequal conformal weights are confined. Since $h_{(\sigma,1)}=1/16$, but $h_{(\sigma,\psi)}=9/16$, the particles $\tilde{\sigma}_1$ and $\tilde{\sigma}_2$ are eliminated from the particle content of the condensed phase. At the third and final step, one verifies that the remaining particles have consistent fusion rules, i.e. the new vacuum is unique and the fusion algebra closes. We find this not to be the case, because
$
(\sigma,\sigma) \times (\sigma,\sigma) = (1,1) + (1,\psi) + (\psi,1) + (\psi,\psi) = 2 \cdot \tilde{1} + 2 \cdot \tilde{\psi},
$
which means that $(\sigma,\sigma)$ must branch into new particle types. The uniqueness of the vacuum can be satisfied if one replaces $(\sigma,\sigma)$ by $\lambda  + \bar{\lambda}$ and demands that these particles satisfy the new fusion rules
\be
\begin{array}{c}
	\tilde{\psi} \times \lambda = \bar{\lambda}, \quad \tilde{\psi} \times \bar{\lambda} = \lambda, \quad\tilde{\psi} \times \tilde{\psi} = \tilde{1}\,, \\
	 \lambda \times \bar{\lambda} = \tilde{1}, \quad \lambda \times \lambda = \bar{\lambda} \times \bar{\lambda} = \tilde{\psi}\,. 
\end{array}
\ee
These are the fusion rules of the $u(1)_4$ theory, which is the theory obtained from the ${\rm Ising}^2$ theory via a condensation transition \cite{bais09}.

\paragraph{Condensation and critical spin chains --}
As the topological quantum numbers of the quasiparticles are in correspondence with the primary fields of the CFT that describes the edge of the topological phase, the condensation mechanism suggests a transition between the critical $H^2_{\rm TFI}$ and $H^1_{\rm XY}$ chains that are described by the ${\rm Ising}^2$ and $u(1)_4$ CFT's, respectively. To induce in $H^2_{\rm TFI}$ the effect of condensing the $(\psi,\psi)$ field,
we constrain the boundary conditions such that the states labeled by the confined particles $\tilde{\sigma}_1$ and $\tilde{\sigma}_2$ are removed from the spectrum. This can be achieved via the boundary term
\be \label{HB}
 H^B_{\rm TFI} = \bigl( \mathcal{P}_o - \id \bigr) \sigma^{x}_{L-2}\sigma^{x}_{0} + \bigl( \mathcal{P}_e - \id \bigr) \sigma^{x}_{L-1}\sigma^{x}_{1}\,,
\ee
that couples the chains on the even and odd sites in a global manner. Namely, the odd parity sector of one chain changes the boundary conditions in the other
(compare to the boundary term in Eq.~\eqref{eq:fermion-tfi}).
As shown in Table~\ref{tab:condensation}, the boundary conditions in both even and odd site chains are forced to be simultaneously either periodic or anti-periodic, which means states labeled by the confined primary fields $\tilde{\sigma}_1$ and $\tilde{\sigma}_2$ are eliminated. It is straightforward to verify that non-local and site-dependent transformation
\begin{align}
\label{eq:transform1}
\sigma^{z}_{2j} &= \tau^{y}_{2j} \tau^{y}_{2j+1} &
\sigma^{z}_{2j+1} &= \tau^{x}_{2j} \tau^{x}_{2j+1} \\
\nonumber
\sigma^{x}_{2j} &= \prod_{i\leq 2j} \tau^{x}_{i} &
\sigma^{x}_{2j+1} &= \prod_{i\geq 2j+1} \tau^{y}_{i} \, ,
\end{align}
inspired by the one used in \cite{f94,jf78,pjp82}  for open chains, gives the explicit microscopic relation between the two TFI chains and the XY chain \cite{fn1}:
\begin{equation}
H_{\rm TFI}^2 + H^B_{\rm TFI} = H_{\rm XY}^1 \ .
\label{eq:XY-TFI-rel}
\end{equation}

\begin{table}[t]
Condensed phase $H_{\rm TFI}^2 + H^B_{\rm TFI} = H_{\rm XY}^1$\\
\begin{tabular}{|c|c|c|c|}
\hline
 $(BC_e,BC_o)$  & ${\rm Ising}^2$ fields & $\mathcal{T}^z$ & $u(1)_4$ fields \\
	\hline
$(1,1)$ & $(1,1)$,$(1,\psi)$,$(\psi,1)$,$(\psi,\psi)$ & 1 & $ 1, \tilde{\psi}$  \\
\hline
$(-1,-1)$ & $(\sigma,\sigma)$ & -1 & $\lambda,\bar{\lambda}$ \\
\hline
$(-1,-1)$ & $(\sigma,\sigma)$ & -1 & $\lambda,\bar{\lambda}$ \\
\hline
$(1,1)$ & $(1,1)$,$(1,\psi)$,$(\psi,1)$,$(\psi,\psi)$ & 1 & $1, \tilde{\psi}$  \\
	 \hline
\end{tabular}
\caption{\label{tab:condensation}
When the boundary term $H^B_{\rm TFI}$ is added to $H^2_{\rm TFI}$, the boundary conditions on each chain depend on the parity sectors of both chains. The allowed boundary conditions of $H_{\rm TFI}^2+H^B_{\rm TFI}$ are thus given by $(BC_e,BC_o)=$($\mathcal{P}_e \mathcal{P}_o $,$\mathcal{P}_o \mathcal{P}_e $), i.e. both chains have to simultaneously have either periodic or anti-periodic boundary. Because ($\mathcal{P}_e$,$\mathcal{P}_o$)=($\mathcal{T}^x$,$\mathcal{T}^y$), the four sectors of $H^2_{\rm TFI}$ with modified boundary conditions map onto the two sectors of $H^1_{XY}$ labeled by $\mathcal{T}^z=\mathcal{T}^x\mathcal{T}^y$.}
\end{table}

\begin{figure}[ht]
		\includegraphics[width=\columnwidth]{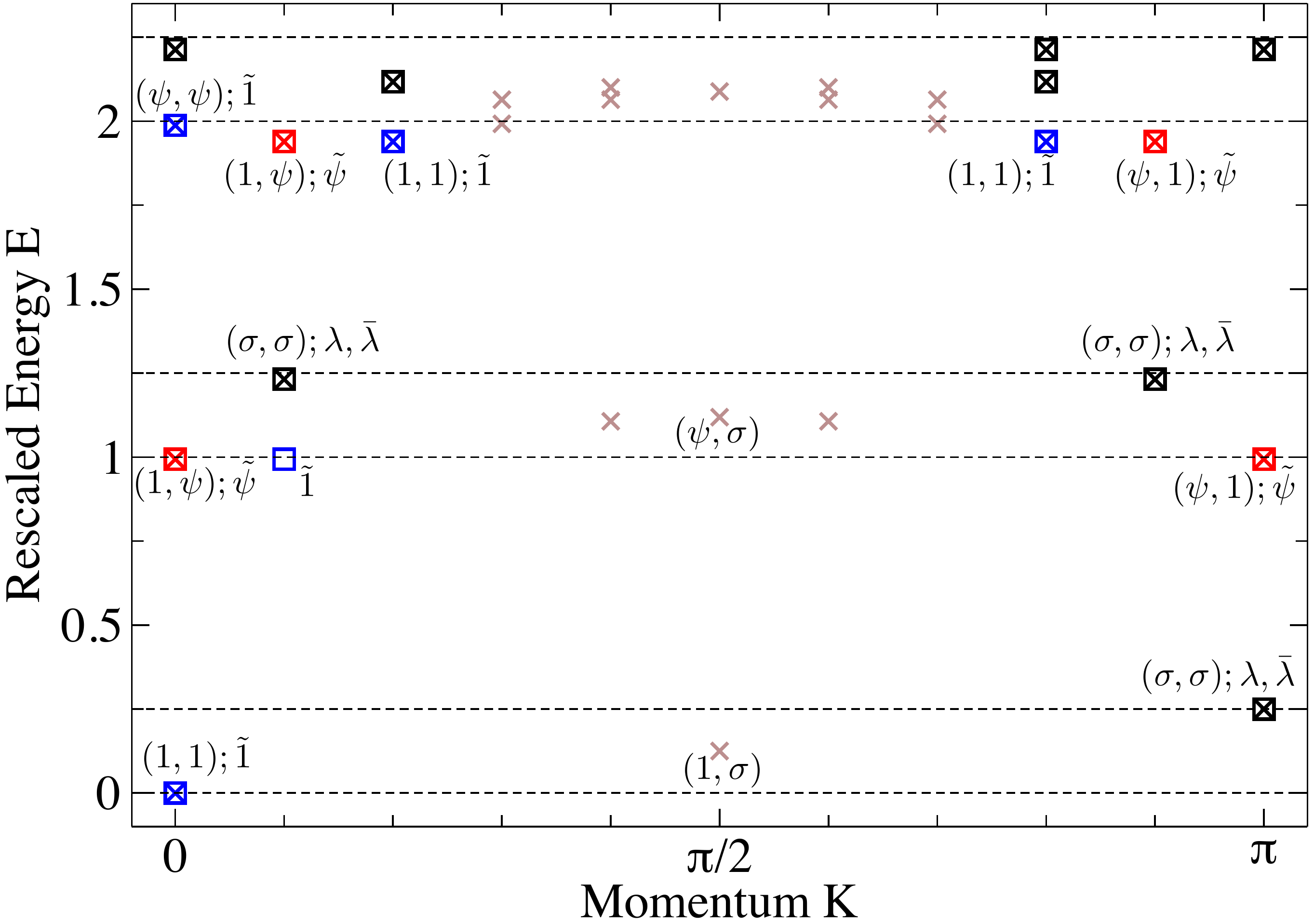}
\caption{\label{spectrum_XY}%
The low-part of the rescaled energy spectra of the $n=2$ TFI chain (crosses) and the $n=1$ XY chain (squares) for $L=20$. The spectra are symmetric under $K\rightarrow -K$. The brown crosses correspond to $H_{\rm TFI}^2$ states labeled by the confined primary fields, whereas the blue, red and black crosses correspond to the states labeled by the $(1,1)$, $(\psi,1)$ and $(\sigma,\sigma)$ fields, respectively. The blue, red and black squares correspond to $H_{\rm XY}^1$ states labeled by the $\tilde{1}$, $\tilde{\psi}$ and $\lambda/\bar{\lambda}$ fields. Whenever the states from $H_{\rm TFI}^2$ and $H_{\rm XY}^1$ coincide, the primary field labeling agrees with the condensation mechanism \cite{fn2}. The dashed lines illustrate the energies predicted by CFT, namely $E= 2h + m$, where $h$ are the scaling dimensions of the primary fields and $m$ is an integer \cite{byb}.
}
\end{figure}

To justify that $H^B_{\rm TFI}$ indeed induces an analogue of a condensation transition, we show that the spectrum encodes also the two other defining features of the mechanism: the identification and the splitting of particles (in addition to the confinement of the $\tilde{\sigma}_1$ and $\tilde{\sigma}_2$ particles).
To show this, we realize that \rf{eq:transform1} gives $\mathcal{P}_e = \mathcal{T}^{y}$ and $\mathcal{P}_o = \mathcal{T}^{x}$, i.e. the sectors of $H_{\rm TFI}^2$ with modified boundary conditions map into the sectors of $H_{\rm XY}^1$, as shown in Table~\ref{tab:condensation}.
Fig.~\ref{spectrum_XY} shows that when the spectra of these models are plotted on top of each other and the individual states are labeled according to the CFT predictions, the condensation picture is confirmed: All $(\psi,\psi)$ states coincide with the new vacuum $\tilde{1}$ states, while all states descending from $(1,\psi)$ and $(\psi,1)$ coincide with the new fermion $\tilde{\psi}$ states; The branching $(\sigma,\sigma) \to \lambda +\bar{\lambda}$ manifests itself as $(\sigma,\sigma)$ states being replaced by two degenerate states labeled by $\lambda$ and $\bar{\lambda}$. 

\paragraph{Deriving critical spin models --}

The condensation mechanism underlying the mapping between the $n=2$ TFI and $n=1$ XY chains suggests that similar mappings should also hold between other critical spin chains whose CFT descriptions can be related by the condensation picture. We employ this insight to derive a critical spin chain described by the $u(1)_2 \times u(1)_2$ CFT.

We start with the critical $n=2$ XY chain, which is described by the
$u(1)_4 \times u(1)_4$ CFT, containing sixteen primary fields (compare to the $n=2$ TFI chain).
Without going through the details of the condensation picture, we note that this
CFT contains a bosonic field, $(\tilde{\psi},\tilde{\psi})$. Condensation of this boson leads
to a pairwise identification of the 16 fields. Four of the
eight resulting pairs turn out to be confined, while none of the fields have to be split.
In the end, one is left with a theory with four fields, whose scaling dimensions and fusion
rules match those of the $u(1)_2 \times u(1)_2$ CFT.

To induce the counterpart of this transition in
$H^2_{\rm XY}$, we again constrain the boundary conditions by introducing the boundary term 
\bq \label{HB_XY}
H^B_{\rm XY} & = & \bigl( \mathcal{T}_o^z - \id \bigr) (\tau^{x}_{L-2}\tau^{x}_{0}+\tau^{y}_{L-2}\tau^{y}_{0}) + \nonumber \\
\ & \ &  \bigl( \mathcal{T}_e^z - \id \bigr) (\tau^{x}_{L-1}\tau^{x}_{1}+\tau^{y}_{L-1}\tau^{y}_{1}),
\eq
which removes the states labeled by the confined primary fields. Using the non-local transformation 
\begin{align} \label{eq:transform2}
\tau^{z}_{2j} &= \rho^{y}_{2j} \rho^{y}_{2j+1} &
\tau^{z}_{2j+1} &= \rho^{x}_{2j} \rho^{x}_{2j+1} \\
\nonumber
\tau^{x}_{2j} &= \bigl( \prod_{i <  2j} \rho^{x}_{i} \bigr)\rho^{z}_{2j} \rho^{y}_{2j+1} &
\tau^{x}_{2j+1} &= \prod_{i \geq  2j+1}  \rho^{y}_{i} \\
\nonumber
\tau^{y}_{2j} &= \prod_{i \leq  2j}  \rho^{x}_{i} &
\tau^{y}_{2j+1} &= \rho^{x}_{2j} \rho^{z}_{2j+1}\bigl( \prod_{i > 2j+1} \rho^{y}_{i} \bigr) \ ,
\end{align}
where the label $j$ takes the values $j=0,1,\ldots,L/2-1$ and
$\rho^\alpha$ are new Pauli matrices,
we arrive at the local and translationally invariant Hamiltonian
$H^2_{\rm XY} + H^B_{\rm XY} = \hat{H}$, where $\hat{H}$ is given by
\begin{align}
\label{HU12}
\hat{H} = 
\sum_{j=0}^{L/2-1} \Bigl(&
\rho^{x}_{2j} \rho^{z}_{2j+1} \rho^{z}_{2j+2} \rho^{x}_{2j+3} +
\rho^{x}_{2j+1} \rho^{x}_{2j+2} \\
+& \rho^{y}_{2j} \rho^{z}_{2j+1} \rho^{z}_{2j+2} \rho^{y}_{2j+3} +
\rho^{y}_{2j+1} \rho^{y}_{2j+2} \Bigr) \ .
\nonumber 
\end{align}
The resulting chain $\hat{H}$ can be solved by means of a Jordan-Wigner
transformation, and we have verified that the critical behavior is indeed described by
the $u(1)_2 \times u(1)_2$ CFT, as predicted by the condensation framework \cite{fn3}.
Although both the starting and resulting CFT's both have a direct product
structure, the condensation mixes the two $u(1)_4$ factors describing
the critical behavior of the $n=2$ XY chain.
We provide more details on the condensation
transition of the $n=2$ XY chain in a forthcoming publication \cite{longversion}.

\paragraph{Conclusions --}
We have shown that condensate-induced transitions between gapped topological phases \cite{bais09} provide a general framework also for transitions between critical spin chains. When the boundary conditions of a spin chain are suitably constrained, some primary fields of the CFT describing the critical behavior are no longer part of the spectrum,  which we interpreted as a precise CFT level counterpart of the confinement of quasiparticles induced by condensation of a boson in a 2D topological phase. Our main result is that such constraining of boundary conditions could be in general achieved through a non-local boundary term. In the presence of such term, we showed that two copies of the critical transverse field Ising chain could be exactly mapped to the critical XY chain, in agreement with the condensation framework. Moreover, we were able to derive a novel local spin chain, with the predicted critical $u(1)_2 \times u(1)_2$ CFT description, from two copies of the critical XY chain.

We expect that the applicability of this framework extends beyond our examples. An exciting prospect is to derive further critical spin chains, with particular interest on ones whose CFT decriptions coincide with the trial wave functions for fractional quantum Hall states. As the entanglement spectra of critical spin chains and topological phases with same boundary CFT description contain intriguing similarities \cite{Thomale10}, critical spin chains may thus offer simple microscopic platform to study complex topological phase transitions. In this respect it would be interesting to study whether the boundary term could be understood more generally in terms of a perturbation in a boundary CFT \cite{oshikawa97}, and what is the counterpart of topological domain walls (see \cite{Bais09-2,gs09,gils09}) in critical spin chains.

\paragraph{Acknowledgements --}

EA thanks T. Quella for stimulating discussions. We would like to acknowledge the financial support by the Swedish science research council (TM) and the Dutch Foundation for Fundamental Research on Matter (VL).

%%%%%%%%%%%%%%%%%%
%% Bibliography
%%%%%%%%%%%%%%%%%%


\begin{thebibliography}{10}
\bibitem{fn1}
This relation had also been studied in \cite{ps84}, although there the boundary term seems to have been missed.
\bibitem{fn2}
As the modified boundary conditions also shift the allowed momenta, not all states have to coincide. To verify the condensation mechanism, it is sufficient that all those states that do coincide are labeled consistent with the predicted identifications \cite{longversion}.
\bibitem{fn3}
We note that upon changing the relative strength of the two- and four-body terms, a gap opens up immediately.
\end{thebibliography}
\end{document}